\documentclass[%
 reprint,
 amsmath,amssymb,
 aps,
]{revtex4-2}

\usepackage{graphicx}% Include figure files
\usepackage{dcolumn}% Align table columns on decimal point
\usepackage{bm}% bold math
\usepackage{xcolor}

\usepackage{times}
\usepackage{hyperref}% add hypertext capabilities
\hypersetup{breaklinks=true}
\newcommand{\QQ}{\mathcal{Q}}
\newcommand{\PP}{\mathcal{P}}
\DeclareMathSymbol{\shortminus}{\mathbin}{AMSa}{"39}

\begin{document}

\preprint{APS/123-QED}

\title{A deep learning approach to the measurement of long-lived memory kernels from generalised Langevin dynamics}

\author{Max Kerr Winter}
\thanks{To whom correspondence should be addressed: m.j.kerr.winter@tue.nl and l.m.c.janssen@tue.nl}
\author{Ilian Pihlajamaa}
\author{Vincent E. Debets}
\author{Liesbeth M. C. Janssen}
\thanks{To whom correspondence should be addressed: m.j.kerr.winter@tue.nl and l.m.c.janssen@tue.nl}
\affiliation{Department of Applied Physics, Eindhoven University of Technology, P.O.~Box 513, 5600 MB Eindhoven, The Netherlands}
\date{\today}

\begin{abstract}
Memory effects are ubiquitous in a wide variety of complex physical phenomena, ranging from glassy dynamics and metamaterials to climate models.  The Generalised Langevin Equation (GLE) provides a rigorous way to describe memory effects via the so-called memory kernel in an integro-differential equation. However, the memory kernel is often unknown, and accurately predicting or measuring it via e.g.\ a numerical inverse Laplace transform remains a herculean task. Here we describe a novel method using deep neural networks (DNNs) to measure memory kernels from dynamical data. As proof-of-principle, we focus on the notoriously long-lived memory effects of glass-forming systems, which have proved a major challenge to existing methods. Specifically, we learn the operator mapping dynamics to memory kernels from a training set generated with the Mode-Coupling Theory (MCT) of hard spheres. Our DNNs are remarkably robust against noise, in contrast to conventional techniques. Furthermore, we demonstrate that a network trained on data generated from analytic theory (hard-sphere MCT) generalises well to data from simulations of a different system (Brownian Weeks-Chandler-Andersen particles). Finally, we train a network on a set of phenomenological kernels and demonstrate its effectiveness in generalising to both unseen phenomenological examples as well as supercooled hard-sphere MCT data. We provide a general pipeline, KernelLearner, for training networks to extract memory kernels from any non-Markovian system described by a GLE. The success of our DNN method applied to noisy glassy systems suggests deep learning can play an important role in the study of dynamical systems with memory.
\end{abstract}

\maketitle

%\tableofcontents

\section{\label{sec:Intro}Introduction}
Non-Markovian systems, i.e.\ those that exhibit memory effects, pose a number of major challenges to both analytic and computational analysis. This issue is of particular importance as such systems occur across many different areas of modern physics, for example climate models \cite{Franzke2015}, gene interaction networks \cite{Herrera-Delgado2018}, quantum-classical simulations \cite{Kelly2016a,Pfalzgraff2019}, and the behaviour of supercooled liquids and glasses \cite{Janssen2018,Debenedetti2001}, among many others. 

A common, and very general, framework for describing non-Markovian dynamics in continuous time is the Generalised Langevin Equation (GLE), where memory effects are included via a so-called memory kernel. GLEs are a common occurrence across statistical physics and beyond as they are produced by the Mori-Zwanzig projection operator formalism \cite{Reichman2005b,Zwanzig1961,Mori1965}. This formalism starts with a memoryless (e.g.\ Hamiltonian) system in a very high dimensional space, and projects the dynamics onto a lower dimensional space consisting of degrees of freedom that are of theoretical interest or are experimentally accessible. The price paid for this dimensionality reduction is the emergence of memory effects in the low dimensional dynamics. Memory kernels are often unknown or highly non-trivial and so measuring them from data is an important step in developing and testing non-Markovian theories. Furthermore, including memory effects can be a highly computationally efficient method of describing complex dynamics \cite{Cao2020}, hence it is desirable to have quick, accurate techniques for measuring memory kernels.

Glasses and supercooled liquids are particularly challenging non-Markovian systems to study as they experience complex dynamics over a wide range of length and time scales, with memory effects lasting multiple orders of magnitude in time. The Mori-Zwanzig method has proved to be very popular in the field of glassy physics, and in particular forms the basis of Mode-Coupling Theory (MCT), a first-principles, self-consistent framework of the glass transition \cite{Bengtzelius1984,Leutheusser1984,Das2004,Gotze2008,Reichman2005b,Janssen2018}. To date, there is no complete theory of the glass transition, and MCT is no exception. Although the Mori-Zwanzig method is exact, it results in an intractable expression for the memory kernel which must then be approximated in a number of ways, varying in complexity depending on the flavour of MCT \cite{Szamel2003,Leutheusser1984,Das1986,Gotze1987,Luo2020,Janssen2015}. This fact emphasises the memory kernel as an object of particular importance to glassy physics, as it is the point at which an exact theory is abandoned in favour of approximations.
%Glasses are a particularly challenging system to study as they experience dynamics over a wide range of length and time scales, with memory effects lasting multiple orders of magnitude in time. 
%In this paper we focus on the glass transition as an archetypal non-Markovian process, with particular emphasis on MCT.

For both glassy and other non-Markovian systems, a GLE of some autocorrelation function, $y$, can be written in the overdamped limit as
\begin{align}
y'(t) + \Omega y(t) +  \int_{-\infty}^t d\tau  K(\tau) y'(t-\tau)  = 0,\label{eq:GLE}
\end{align}
where $K$ is the memory kernel, and $\Omega$, the so-called frequency term, is a parameter that is in general known, which describes the memoryless evolution of $y$. In general GLEs also contain a random force term, which is removed from Eq.~\ref{eq:GLE} by taking the correlation of a variable with itself to get the autocorrelation $y$. In MCT, $y$ is the autocorrelation function of density fluctuations. Equations with the form of Eq.~\ref{eq:GLE} are also called memory equations. The question of how to study $K$ given $y$ at first appears simple. By applying a Laplace transform to Eq.~\ref{eq:GLE} the kernel can be disentangled from its convolution with $y'$, resulting in an explicit expression for $K$,
\begin{align}
K(t) = \mathcal{L}^{\shortminus1}\left [ \frac{y(0) - (s + \Omega) \mathcal{L}[y]}{s \mathcal{L}[y]-y(0)}\right ] \label{eq:K},
\end{align}  
where $\mathcal{L}[f(t)](s) = \int_0^\infty e^{-s t} f(t) dt$ is the Laplace transform of a function, $f(t)$. The complexity arises from the fact that in practice performing an inverse Laplace transform is a major challenge. 

The difficulties of performing a numerical inverse Laplace transform are well known \cite{McWhirter1978,Craig1994,Satija2019}, and the extent of this problem is nicely summarised by Epstein and Schotland: ``Our results give cogent reasons for the general sense of dread most mathematicians feel about inverting the Laplace transform'' \cite{Epstein2008}. Laplace inversion is an example of an ill-posed inverse problem, where information is lost during any numerical implementation of the forward transform, making the inverse difficult if not impossible. This effect can be demonstrated by considering the forward Laplace transform of a function expressed as a Fourier series, $\mathcal{L}[f(t)] = \mathcal{L}[\sum_{i} a_i \sin(\omega_i t) + \sum_{j} b_j \cos(\omega_j t)]$. The Laplace transforms of $\sin$ and $\cos$ are $\mathcal{L}[\sin(\omega t)](\tau) = \omega/(\omega^2 + \tau^2)$, and $\mathcal{L}[\cos(\omega t)](\tau) = \tau/(\omega^2 + \tau^2)$ respectively. As such, the amplitude of high-frequency components in $f(t)$ are suppressed by the $\omega^2$ in the denominators of the respective Laplace transforms. Above some critical $\omega^\ast$, the high-frequency components become indistinguishable from noise (either experimental or numerical) and hence are unrecoverable by the inverse transform. Consequently, numerical inverse Laplace transforms are very sensitive to noise, with even numerical round-off errors potentially overwhelming the signal. 

Many authors in the soft matter and glassy physics communities have taken an alternative approach to determining the memory kernel, whereby an implicit Volterra integral equation for $K$ is constructed from pair-wise correlation functions \cite{Baity-Jesi2019a,Vroylandt2022,Han2021,Cao2020,Obliger2023}. Such correlation functions are constructed by taking an ensemble average over a large number of trajectories of particle-based simulations. A variety of different numerical methods have been used to successfully solve such equations for the short time memory effects of various systems \cite{Jung2017,Kowalik2019,Berkowitz1981,Shi2004}.

An alternative to both explicit Laplace inversion, and solving a Volterra equation, is to use a minimisation approach to approximate the kernel. The problem of performing the inverse Laplace transform in Eq.~\ref{eq:K} can be reformulated as finding some $\bar{K}$ such that $|\mathcal{L}[\bar{K}] - \mathcal{L}[K]|<\epsilon$, for some $\epsilon\in\mathbb{R}$ that can be set arbitrarily small  \cite{Craig1994}. As the high-frequency parts of $\bar{K}$ and $K$ are lost in the forward transform, this minimisation problem does not have a unique solution. Consequently, it is helpful to include a regularisation functional, $\mathcal{R}$, resulting in the minimisation problem $|\mathcal{L}[\bar{K}]-\mathcal{L}[K]| + |\lambda \mathcal{R}[\bar{K}]|<\epsilon$, where $\lambda$ is a parameter to control the strength of the regularisation. A simple regularisation technique is to penalise low order derivatives in the solution, i.e.\ $\mathcal{R}[\bar{K}] = \bar{K}''$, resulting in a smooth $\bar{K}$. In some cases, e.g.\ astronomical image restoration, an entropy can be calculated for the image function, which is then maximised by a regularisation functional \cite{Narayan1986}. 
Recent successes in modelling systems with memory effects over a wide range of timescales have also been achieved by moving away from the GLE completely in favour of time-local non-Markovian methods with attractive computational properties \cite{Dominic2023}.

Complementary to conventional physical and mathematical approaches, machine learning techniques are rapidly emerging as powerful and computationally efficient tools for the study of glasses \cite{Schoenholz2016,Tah2022,Alkemade2022,Coslovich2022,Paret2020,Bapst2020,Shiba2023,Boattini2020,Pezzicoli2022,Jung2022} and complex soft matter systems more generally \cite{Campos-Villalobos2021,She2023}. As such, the recent explosion of interest in deep learning suggests a new route to extracting memory kernels from GLEs following the minimisation philosophy above. Deep Neural Networks (DNNs) have very attractive generalisation and expressivity properties \cite{Smith2020,Park2019,Hornik1989}, and so it is reasonable to ask whether a DNN could learn the mapping from the function $y$ to the kernel $K$ in Eq.~\ref{eq:GLE}. Previous authors have made significant progress by parameterising the kernel with manually curated sets of functions \cite{Lei2016,Grogan2020,Ayaz2021}. Our neural network parameterisation follows a similar philosophy while taking advantage of the very broad approximation properties of DNNs. Such an approach has shown impressive results when applied to analytic problems with rapidly decaying memory kernels \cite{Fournier2020}. By learning a mapping between function spaces, our approach falls within the rapidly expanding field of deep operator learning, which was popularised by the publication of DeepONet \cite{Lu2021a}.

In this work we present a novel machine-learning based method for kernel extraction from GLE data. As a demonstration of effectiveness, we train and apply DNNs to the challenging problem of extracting memory kernels in glass-forming systems that have very long-lasting memory effects and significant levels of noise. The networks are far more robust to noise in the input signal than traditional methods. Furthermore, we show that a network trained on data generated from analytic theory (hard-sphere MCT) generalises well to data from particle-based simulations of a different system. MCT provides an analytic, though approximate, memory kernel from which a training set can be generated, whereas with simulations we study the true dynamics, but with an unknown kernel. Our network that has been trained on MCT data is available at \url{https://zenodo.org/record/7603275#.Y_vAxS8w1pQ}.

While we focus on glass-forming materials here to establish proof of principle, our method is not limited to glassy systems per se. In fact we hope it will be used to study the form of kernels across a wide variety of non-Markovian phenomena, and in particular systems that have a less well developed body of theory than glassy materials. Furthermore, memory effects play an important role in coarse-grained or reduced order models (ROM) of complex systems \cite{Gouasmi2017,Ahmed2019,Li2015}. As such, being able to measure kernels from a minimum of high resolution data is important. Consequently, we have written a pipeline for users to train networks on their GLE system of choice, which is available at \url{https://github.com/mkerrwinter/KernelLearner}. 

\section{\label{sec:setup} Problem definition and numerical setup}
Our general method can be briefly summarised as follows. %An overview of our method is the following. 
Starting from a set of memory kernels similar to those we wish to measure, training and testing sets are generated by solving the GLE using the same method as in \cite{Debets2021a}, subjecting the solutions to many noise realisations, and using the (solution, kernel) pairs as input and output respectively. Multiple networks are trained over a range of hyperparameters, and an optimum network is selected which achieves the minimum test loss. Finally, a novel memory kernel can be measured from unseen input data, and validation can be performed by solving the GLE with this measured kernel to compare with the input. The ``first guess'' kernels could be derived from theory, generated to a low level of accuracy with existing kernel measurement methods, measured from a similar system, or simply be informed guesswork.

Our goal is to extract a memory kernel, $K(t)$, from the density autocorrelation function, $F(t)$, of a glassy system described by a GLE like Eq.~\ref{eq:GLE}. We use $F(t)$ curves generated by numerically solving MCT, as well as curves measured from particle-based simulations. The neural networks are trained on MCT data, and validation is performed on both unseen MCT data and simulation data. For comparison, we also extract the kernel by applying conventional (i.e.\ non-network) methods.

As the variable of interest, MCT typically employs the autocorrelation function
\begin{align}
F(k, t) = \frac{1}{N} \langle \rho_{-\textbf{k}}(0)\rho_{\textbf{k}}(t) \rangle,
\end{align}
where $\rho_\textbf{k}=\sum_j e^{i\mathbf{k}\cdot\mathbf{r}_j(t)}$ is the microscopic density in Fourier space, $\textbf{k}$ a wavevector, $N$ the number of particles, and $\langle\cdot\rangle$ denotes an ensemble average. The overdamped MCT GLE has the form
\begin{align}
\dot{F}(k, t) + \frac{D_0k^2}{S(k)}F(k, t) + &\nonumber \\ \int_0^t d\tau K_{\text{MCT}}&(k, \tau)\dot{F}(k, t-\tau) = 0 \label{eq:MCT},
\end{align}
where $k=|\textbf{k}|$ is the wavenumber, $D_0$ is the self-diffusion coefficient, $S(k) = F(k, 0)$ is called the static structure factor, and $K_{\text{MCT}}$ is the kernel. The MCT kernel is given by
\begin{align}
K_{\text{MCT}}(k, t) = \frac{\rho D_0}{16\pi^3} \int d\textbf{q}  V_{\textbf{q}, \textbf{k}-\textbf{q}}^2 F(q, t) F(|\textbf{k}-\textbf{q}|, t),
\end{align}
where $\rho$ is the average number density, the vertex term $V_{\textbf{q}, \textbf{k}-\textbf{q}} = k^{-1}[\textbf{k}\cdot \textbf{q}c(q) + \textbf{k}\cdot(\textbf{k}-\textbf{q})c(|\textbf{k}-\textbf{q}|)]$, and $c(k)=\rho^{-1}[1-1/S(k)]$ (for more details see e.g.\ \cite{Janssen2018}). In general, $F(t)$ is a function of the wavenumber, and the MCT kernel couples different wavenumbers together. Equation~\ref{eq:MCT} is subject to the initial condition $S(k) = F(k,0)$. 

To numerically solve the MCT equation we use the Percus-Yevick closure for a system of hard spheres \cite{Percus1958,Wertheim1963}. This is an analytic approximation for the static two-point correlation function $S(k)$ of a system of hard spheres at a given density. For simplicity, we test our method on the $F(k,t)$ behaviour at wavenumber $k=k^\ast$, corresponding to the main peak of $S(k)$. From here on we will omit the explicit $k$ dependence and use $F(t)=F(k^\ast, t)$ for brevity. Note that by restricting ourselves to the peak wavenumber we only achieve accurate network predictions at $k^\ast$, however the training set can easily be extended to include more wavenumbers.

The simulation data we use is from a system of particles interacting via the Weeks-Chandler-Andersen (WCA) potential \cite{Weeks1971}. Details of the simulations are given in Appendix \ref{sec:appendix_A}. Note that the type of particle in the simulation data (soft spheres) is different from the MCT data used to train the networks (hard spheres).

\subsection{Dataset, network and training process}
\begin{figure}[h]
\includegraphics[width=9cm]{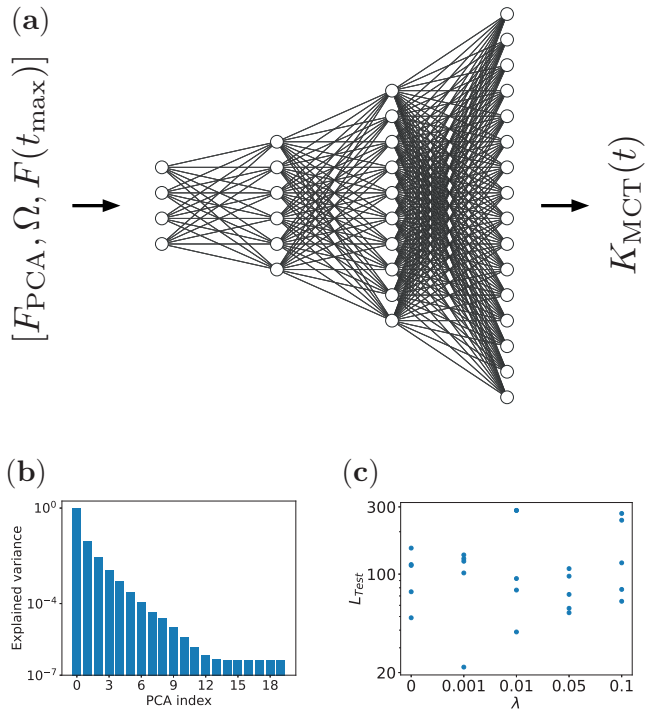}
\caption{$\mathbf{(a)}$ A schematic of the network structure. The input consists of the first 15 PCA components of a pairwise correlation function $F(t)$, the frequency term $\Omega$, and the final correlation value $F(t_{\text{max}})$. The output is an approximation of the memory kernel used to generate $F(t)$ via Eq.~\ref{eq:MCT} \cite{NN-SVG}. $\mathbf{(b)}$ The explained variance of consecutive PCA components of the training set, which display a kink between components 12 and 15. $\mathbf{(c)}$ Minimum test loss values for a subset of networks in the hyperparameter search. These networks have a width factor $\omega=8$, and were trained with a batch size of 2500. The plot shows the variability between networks with different regularisation $\lambda$, and different initial conditions.}
\label{fig:network}
\end{figure}

Both the training dataset and MCT testing dataset are produced in the same way. First, Eq.~\ref{eq:MCT} is solved numerically at volume fractions $\phi \in \{0.45, 0.451, 0.452..., 0.58 \}$, that are symmetric about the MCT glass transition $\phi_g \approx 0.516$ \cite{Bengtzelius1984}, to produce a set of analytic curves denoted $F_A(t)$. Each curve is then subjected to 1000 different realisations of Gaussian noise like
\begin{align}
F(t) = F_A(t) + \mu (\text{max}(F_A)-\text{min}(F_A)) \xi(t),\label{eq:noise_def}
\end{align} 
where $\xi(t)$ is Gaussian noise with unit variance and zero mean, and $\langle \xi(t) \xi(t')\rangle = \delta(t-t')$. The parameter $\mu$ controls the strength of the noise. By adding noise to the training set the DNN learns the mapping from noisy dynamics to a clean memory kernel, and hence is able to handle real data measured from simulations or experiments.  The dimension of the set of noisy curves, $\{ F(t)\}$, is then reduced with Principal Component Analysis (PCA) where only the first 15 PCA components are retained \cite{Jolliffe2002}. Note that PCA is a generic and automated method for finding a suitable lower-dimensional representation of data, i.e.\ it does not rely on any domain specific knowledge about how best to encode the data, and so allows this pre-processing method to be broadly applicable to many different systems. Concretely, our original dataset consisting of $P$ examples, each defined on a time grid of 4352 points (with the grid spacing doubling as in \cite{Janssen2015}), is transformed to $P$ examples in a 15-D space which serves as our DNN-input. 
%It is always desirable to perform machine learning in as low dimensional a space as possible without losing important information. PCA is a generic and automated method for finding a suitable subspace, i.e. it does not rely on any domain specific knowledge about how best to encode the data, and so allows our method to be broadly applicable to many different systems. 
The explained variance per principal component is shown in Fig.\ \ref{fig:network}$\mathbf{(b)}$, which levels off at the 15th component. As well as the PCA components, we include the frequency term, $\Omega=\frac{D_0k^2}{S(k)}$, and the value of the autocorrelation function at the final timepoint, $F(t_{\text{max}})$, in order to clearly distinguish between liquids ($F(t_{\text{max}})=0$) and glasses ($F(t_{\text{max}})>0$). The dimensionally reduced set of noisy curves, $\Omega$, and $F(t_{\text{max}})$ form the input of the network, of dimension 17. The target is the $K_{\text{MCT}}$ used to produce a given $F(t)$, discretised on a logarithmically spaced time grid of 100 points. Consequently, the output of the neural network is also defined on this time grid. Both the testing and the training set consist of $P=65500$ such examples.

We use a fully connected feed-forward network with $L+1$ layers, with input $\mathbf{x}$, and output given by
\begin{align}
f_\alpha(\mathbf{x}, \boldsymbol{\Theta}) &= a_\alpha^{L+1},\\
a_\beta^i &= \sum_\alpha W_{\alpha \beta}^i \sigma(a_\alpha^{i-1}) - B_\beta^i,\qquad 2 \leq i \leq L+1\\
a_\beta^1 &= \sum_\alpha W_{\alpha \beta}^1 x_\alpha - B_\beta^1.\label{eq:net_def}
\end{align} 
Greek letters index the matrices of weights between each layer, whereas Latin letters index the layers themselves, hence $W_{\alpha \beta}^i$ are the elements of the matrix of weights between layer $i-1$ and $i$. The biases are given by $B_\alpha^i$. The variable $\boldsymbol{\Theta}$ is a vector containing all weights and biases. For the activation functions we choose the popular Rectified Linear Unit (ReLU), $\sigma(z) = \text{max}(0, z)$. We use the dropout method to reduce overfitting \cite{Srivastava2014}, so that at each training step nodes are temporarily dropped from the network with probability $p=0.5$. The network structure consists of $L=6$ hidden layers that gradually increase in width. The hidden layer widths are $[50\omega, 100\omega, 150\omega, 200\omega, 250\omega, 300\omega]$, where the width factor $\omega$ is an integer hyperparameter controlling the width of the network. We have borrowed this triangular network structure from the authors of \cite{Fournier2020}, who address a similar problem. We chose a fully connected network (as opposed to e.g. a convolutional neural network) as it is the most general feed forward architecture, and is determined by a relatively small set of hyperparameters. The structure of the network, input, and output data is shown schematically in Fig.\ \ref{fig:network}$\mathbf{(a)}$.

We use a weighted mean square error loss function between network output at neuron $j$, $f_j$, and the true kernel in adimensional form $\tilde{K}(t_j)$, with L2 regularisation on the network parameters, $\Theta_k$, to prevent overfitting \cite{Ying2019},
\begin{align}
L = \frac{1}{P}\sum_{i=1}^P \frac{1}{j_{\text{max}}} \sum_{j=1}^{j_{\text{max}}}\alpha_j(f_j-\tilde{K}(t_j))^2 + \lambda \sum_{k=1}^M \Theta_k^2,\label{eq:loss_func}
\end{align}
where the first sum is over examples in the training set, and the second is over the time points $t_j$ at which $\tilde{K}$ is discretised. The training set is produced using natural units. Space is measured in terms of the particle diameter, $d$, and time is measured in $d^{2}/D$, where $D$ is the diffusion constant. The units are chosen such that $d=D=1$, and $\tilde{K}=K d^2/D$. The weights increase linearly with the time grid, $\alpha_j = j/j_{\text{max}}$, so that long-time behaviour is given more importance. This is because the kernel at long times affects the dynamics to a greater extent than at short times. The parameter $\lambda$ controls the strength of L2 regularisation over the $M$ network parameters. Its effect on the test loss is demonstrated in Appendix \ref{sec:appendix_overfitting}. Training is performed using the Adam method, a popular stochastic gradient descent algorithm \cite{Kingma2015}. We use early stopping to avoid overfitting, i.e.\ we select the optimum network state that gives the minimum test loss, as demonstrated in Appendix \ref{sec:appendix_overfitting}. 

\subsection{Non-network methods}
For comparison with the above network-based method, we also implement two traditional kernel extraction methods. The first applies an inverse Laplace transform to Eq.~\ref{eq:K}. We use the De Hoog algorithm to evaluate $\mathcal{L}^{-1}$ using a Fourier series with accelerated convergence \cite{DeHoog1982,Kuhlman2013}. This method outperformed other common inversion algorithms (Talbot \cite{Talbot1979} and Stehfest \cite{Stehfest1970}) on our data. To mitigate the effect of noise we first smooth the $F(t)$ curves with a Savitzky-Golay filter \cite{Savitzky1964} before applying the De Hoog inverter. The second method is to construct and solve a Volterra integral equation for $K$, the details of which are given in Appendix \ref{sec:volterra_appendix}. 

\section{\label{sec:numerics} Results}
\subsection{Hyperparameter search}
We first search for optimum network hyperparameters. As there are a large number of hyperparameters that can be optimised a full grid search is unfeasible. Instead, we choose the L2 regularisation strength $\lambda$, the batch size of the Adam method, and the width factor $\omega$ as the most important hyperparameters, and perform a search over reasonable intervals for each. This process is shown for $\lambda$ in Fig.\ \ref{fig:network}$\mathbf{(c)}$, where multiple initial conditions have been included for each $\lambda$. As well as demonstrating the effect of $\lambda$ on the test loss, this figure illustrates the significant randomness introduced by using different initial conditions for $\boldsymbol{\Theta}$, and hence the importance of training multiple networks with different initial conditions. We select the network with the lowest test loss across the whole hyperparameter search.

\subsection{Performance on MCT Percus-Yevick hard spheres}
We apply our optimum network to the task of extracting memory kernels from unseen $F(t)$ curves generated by hard sphere MCT and subjected to noise according to Eq.~\ref{eq:noise_def}. For comparison, we use the De Hoog algorithm to extract the kernel by means of Eq.~\ref{eq:K} applied to noisy $F(t)$ curves. It is reasonable to ask whether a simple smoothing procedure to mitigate the effect of noise would be sufficient to achieve reasonable performance without resorting to deep learning. To investigate this, we also use the De Hoog algorithm on $F(t)$ curves that have been smoothed by a Savitzky-Golay filter.

\begin{figure}[h]
\includegraphics[width=8.7cm]{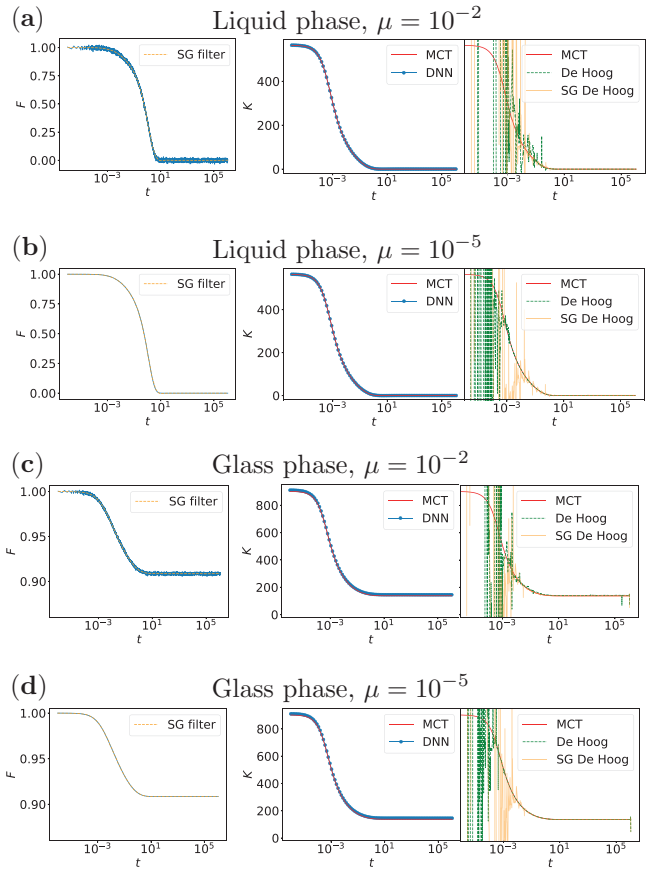}
\caption{Examples of kernel extraction from a density autocorrelation function using a trained network and the De Hoog algorithm. $\mathbf{(a)}$ The liquid phase of Percus-Yevick Hard Sphere MCT with volume fraction $\phi=0.475$. The left-hand panel shows the noisy $F(t)$ curve, with $\mu=10^{-2}$ as defined in Eq.~\ref{eq:noise_def}, as well as the smoothed curve produced by the Savitzky-Golay filter. The middle panel contains the memory kernel as measured by a network and the true MCT kernel. The right-hand panel contains the kernel measured by the De Hoog algorithm from the smoothed $F(t)$ (`SG De Hoog'), the kernel measured from the raw noisy curve (`De Hoog'), and the true MCT kernel. $\mathbf{(b)}$ The same curves measured from $F(t)$ at $\phi=0.475$ with $\mu=10^{-5}$. $\mathbf{(c)}$ and $\mathbf{(d)}$ are the same measurements again but in the glass phase with $\phi=0.52$.}
\label{fig:MCT_results}
\end{figure}

The left-hand panels of Fig.\ \ref{fig:MCT_results} show both the noisy and smoothed $F(t)$. The two right-hand panels show the kernel measured by our neural network, Laplace inversion of Eq.~\ref{eq:K}, and Laplace inversion of Eq.~\ref{eq:K} using the smoothed $F(t)$. We measure kernels in both the liquid (volume fraction $\phi=0.475$) and glass ($\phi=0.52$) regimes, subjected to both high ($\mu=10^{-2}$) and low ($\mu=10^{-5}$) levels of noise. In all cases the network reproduces the true MCT kernel from $F(t)$ to a very high degree of accuracy. In contrast, Laplace inversion fails to produce an accurate (or indeed even vaguely reasonable) kernel across all times. Although the smoothing process on $F(t)$ in the left-hand panel of Fig.\ \ref{fig:MCT_results}$\mathbf{(a)}$ and $\mathbf{(c)}$ looks very effective by eye, it does not significantly improve the accuracy of the measured kernel. As can be seen in the low noise plots, Fig.\ \ref{fig:MCT_results}$\mathbf{(b)}$ and $\mathbf{(d)}$, the noise on $F(t)$ must be reduced to the point where it is no longer visible by eye before conventional Laplace inversion can measure a reasonable kernel over multiple decades. Even then, this approach fails at short times. Interestingly, when noise is this low the smoothing process actually decreases performance. 

\begin{figure}[h]
\includegraphics[width=8.6cm]{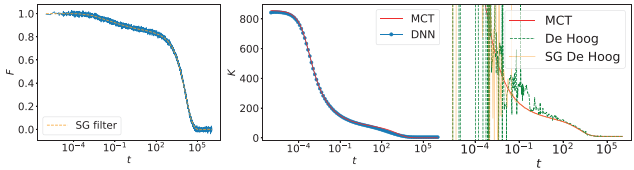}
\caption{Left panel: density correlation function $F(t)$ in the supercooled regime as predicted by MCT for hard spheres at volume fraction $\phi=0.515$, which is very close to the MCT glass transition $0.515<\phi_{g}<0.516$. Middle panel: the corresponding MCT memory kernel and the kernel predicted by DNN. Right panel: the corresponding memory kernel as obtained by the De Hoog algorithm, applied to both noisy and smoothed $F(t)$ data.  
%Kernels extracted from $F(t)$ (left-hand panel) by DNN (middle panel) and by the De Hoog algorithm applied to both noisy and smoothed data (right-hand panel). Here, the volume fraction $\phi=0.515$ is very close to the MCT glass transition volume fraction which lies in $0.515<\phi_{g}<0.516$.
}
\label{fig:phi_near_glass}
\end{figure}

In Fig.\ \ref{fig:phi_near_glass} we demonstrate that the DNN hugely outperforms De Hoog Laplace inversion once again, this time in the supercooled regime, very close to the glass transition point. This is a particularly challenging region of parameter space as $F(t)$ becomes very sensitive to small changes in the initial conditions. Fig.\ \ref{fig:novel_phis} in Appendix \ref{sec:appendix_B} shows how the network extrapolates well to regions of phase space not included in the training set. The ability to generalise is crucial for the usefulness of the DNN method. The caveat to this extrapolation is that deep in the glass phase, at volume fractions higher than those in the training set, the dynamics become only weakly dependent on the kernel and the performance of the network decreases.

The results in Fig.\ \ref{fig:MCT_results} demonstrate the extreme susceptibility of Laplace inversion to noise, as well as the huge improvement achieved by deep learning. Furthermore, as shown in Fig.\ \ref{fig:phi_near_glass} and Appendix \ref{sec:appendix_B}, our network method is highly effective in difficult regions of parameter space, and in regions not included in the training set. Finally, the trained DNN is hundreds of times faster than the De Hoog algorithm. Our deep learning method comprehensively outperforms conventional Laplace inversion for measuring the memory kernel of hard sphere MCT.

\subsection{Performance on simulated soft spheres}
To test the limits of the network's performance, we apply it to data from a different system than the one it was trained on. We run Brownian dynamics simulations of WCA monodisperse spheres (see Appendix \ref{sec:appendix_A}), and measure $F(t)$ both from single trajectories, and an ensemble average of several hundred trajectories. These simulations are in the liquid regime, at temperatures just above the solid-liquid binodal below which the system crystallises. We measure kernels by two methods, namely from our neural network, and by constructing an implicit Volterra integral equation for $K$ (see Appendix \ref{sec:volterra_appendix}). The neural network has not been retrained, i.e.\ it has only seen the MCT hard sphere training set of the previous section. This is in order to study the ability of the network to generalise to an unseen system. For validation, we solve Eq.~\ref{eq:MCT} using the kernels measured by the network or Volterra method in place of $K_{\text{MCT}}$, resulting in a new $F(t)$ curve which we can compare with $F(t)$ from the simulations.
\begin{figure}[h]
\includegraphics[width=9cm]{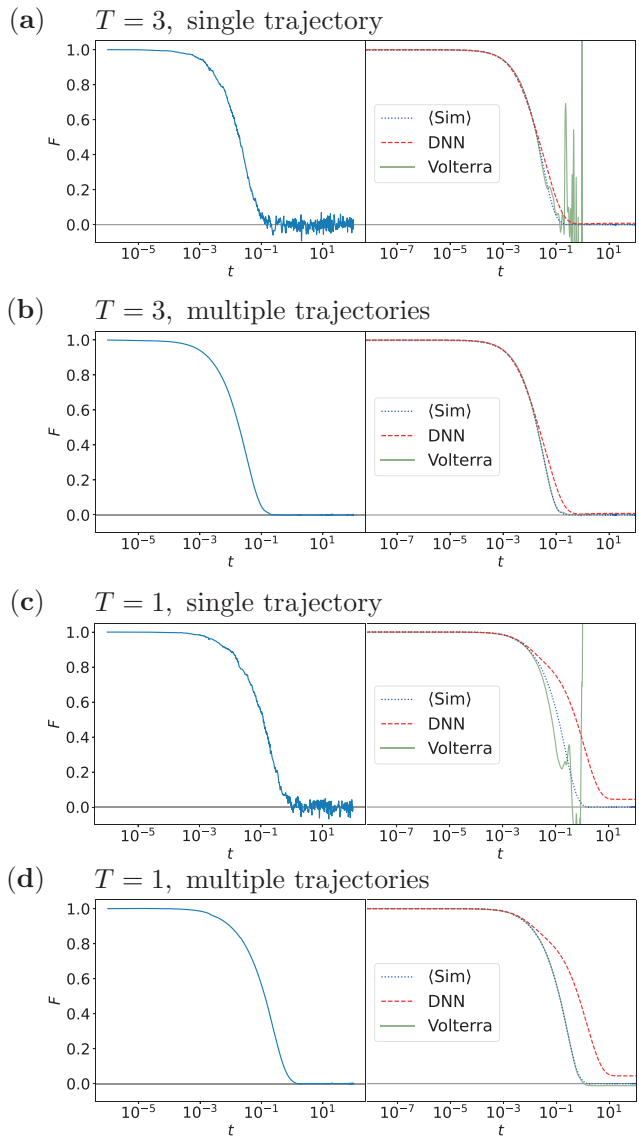}
\caption{Reproducing $F(t)$ from WCA simulations. In the left-hand panels of $\mathbf{(a)}$ and $\mathbf{(c)}$, $F(t)$ is calculated from a single simulation trajectory and hence exhibits significant levels of noise. In the left-hand panels of $\mathbf{(b)}$ and $\mathbf{(d)}$ and in the right-hand panels clean $F(t)$ curves are produced by averaging over many trajectories (`$\langle$Sim$\rangle$'). Kernels are measured from $F(t)$ with the network (`DNN') and Volterra (`Volterra') methods, which are then used to solve Eq.~\ref{eq:MCT}. The resulting $F(t)$ curves are plotted in the right-hand panels. A perfect kernel measurement would result in exactly the same curve as `$\langle$Sim$\rangle$'. In the low-noise plots the Volterra line is almost indistinguishable from the data. Temperatures are in Lennard-Jones units, defined in Appendix \ref{sec:appendix_A}.}
\label{fig:WCA_results}
\end{figure}
In Fig.\ \ref{fig:WCA_results}$\mathbf{(a)}$ the noisy $F(t)$ of a single simulation trajectory (left-hand panel) is used as input. The network is able to reproduce dynamics that are very close to the true dynamics of the system calculated by averaging many trajectories (right-hand panel). In contrast, the Volterra method, which is at first more accurate than the DNN, begins to oscillate wildly and soon diverges. The same measurements are repeated at a lower temperature (but still in the normal liquid regime) in Fig.\ \ref{fig:WCA_results}$\mathbf{(c)}$. Similar to the hard-sphere case, the DNN method hugely outperforms traditional methods in the presence of noise. In Fig.\ \ref{fig:WCA_results}$\mathbf{(b)}$ and $\mathbf{(d)}$ the input data is the ensemble-averaged $F(t)$. The right-hand panels compare the dynamics using the network kernel, and the Volterra kernel, to the true dynamics from the left-hand panel. In this low-noise context the Volterra method performs very well. At high $T$ the network is able to produce dynamics that are very similar to the ground truth, whereas at low $T$ the network is less accurate, but not catastrophically so. This discrepancy is due to the greater role Markovian dynamics (controlled by the frequency term, $\Omega$) play at high temperatures. An explicit comparison between the simulated dynamics, network prediction, and MCT is given in Appendix \ref{sec:WCA_comparison}. Importantly, the network-predicted dynamics exhibit an incorrect non-zero plateau at long times. The long-time behaviour of $F(t)$ is very sensitive to the long-time behaviour of $K$, hence small non-zero values in the tail of $K$ (a likely outcome of any minimisation routine) can result in a non-zero tail in $F(t)$. 

Let us now discuss the computational efficiency of the DNN approach as compared to that of the Volterra method. After generating the HS-MCT training set (which took approximately 35 minutes on a 2020 Macbook Pro),  
the DNN was trained to its minimum test loss state on a single Xeon E5 (2019) CPU core in 24 hours. Measuring a kernel with the trained network subsequently takes less than a second. The dynamics shown in Fig.\ \ref{fig:WCA_results}$\mathbf{(b)}$ and \ref{fig:WCA_results}$\mathbf{(d)}$ (Volterra curves) each required approximately 100 hours of computing time, again on a single Xeon E5 core. As such, the DNN method is significantly faster in this instance. However, a note of caution is required. Many factors can dramatically effect the efficiency of the DNN method, both slowing it down (performing hyperparameter searches, using more data, using larger networks) and speeding it up (using a learning rate schedule, using GPUs,  transfer learning), and it is important to note that minimising computing time was not a priority in this work. Similarly for the Volterra method, simulating different systems, of different sizes, and for different amounts of time, will have a large impact on the efficiency. 

As well as demonstrating that the DNN can accurately measure kernels in a system that differs from the training set, Fig.\ \ref{fig:WCA_results} is also an example of how kernel extraction can be used in reduced order modelling. The `DNN' curves are generated by the relatively cheap process of training a network on HS-MCT theory, measuring the kernel from a single simulation trajectory, then solving a GLE with this kernel. The resulting dynamics closely approximate the much more expensive `$\langle \text{Sim} \rangle$' curves, which were generated by running many repeats of the full particle resolved simulations.

It is important to recall that the DNN has not been retrained on the soft sphere simulations. The input data in Fig.\ \ref{fig:WCA_results} differs from the training set in how it was generated (simulation vs theory), the nature of the noise on $F(t)$, and in the system itself (WCA vs hard spheres). Despite these multiple differences, our results demonstrate how well our DNN method generalises to new systems. Furthermore, the network can reproduce ensemble-averaged dynamics from noisy data, unlike conventional methods, allowing a clean measurement to be made from a single simulation trajectory.

\subsection{A phenomenological training set}
In many situations there is no established theory from which to construct a training set of GLE solutions and memory kernels. In such a case it is instead possible to use physical intuition and educated guesswork to generate training data. As a proof of principle, we demonstrate this process on a training set of phenomenological kernels that exhibit liquid-like, supercooled liquid-like, and glass-like behaviour, i.e.\ simple fast relaxation, two step relaxation, and persistent memory effects respectively. 

\begin{figure}[h]
\includegraphics[width=8.8cm]{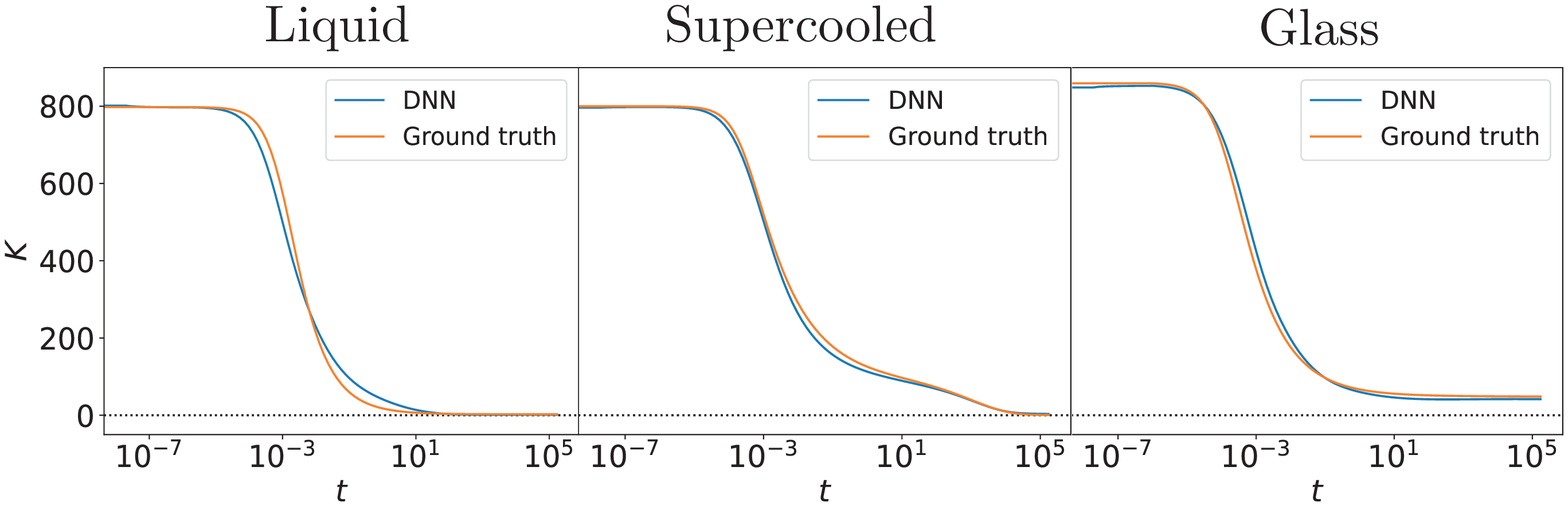}
\caption{Kernel predictions made by the DNN trained on phenomenological kernels when applied to unseen, noisy $F(t)$ in the liquid-like, supercooled liquid-like and glass-like regimes.}
\label{fig:phenom_results}
\end{figure}

Our phenomenological memory kernels are parameterised as
\begin{align}
K(t) = \frac{a}{(1+b t^c)^d} + f e^{-(\frac{t}{10^g})^h},
\label{eq:phenom_K}
\end{align}
where the parameters $a$, $b$, etc. are chosen manually to mimic the behaviour of the three regimes (specific values are given in Appendix \ref{sec:phenom_appendix}). The form of Eq.~\ref{eq:phenom_K} was chosen to capture both fast and slow relaxation regimes. Furthermore, it is known that the approach to the supercooled plateau is a power law, and the long time relaxation can be well fitted to a stretched exponential \cite{Ediger1996,Reichman2005b,Janssen2018}. The first and second terms of $K$ are chosen to reflect this. Next, we solve the GLE using these memory kernels to obtain corresponding $F(t)$ curves, add noise using the procedure in Eq.~\ref{eq:noise_def} with $\mu=10^{-2}$, and construct a training set of (solution, kernel) pairs. The performance of the network on unseen, noisy $F(t)$ curves generated from kernels parameterised by Eq.~\ref{eq:phenom_K} is shown in Fig.\ \ref{fig:phenom_results} for liquid-like, supercooled liquid-like and glass-like kernels. The network predictions in all three regimes are accurate, and hence demonstrate the ability of the network to learn a training set of kernels with the complex parameterisation of Eq.~\ref{eq:phenom_K}.

\begin{figure}[h]
\includegraphics[width=8.8cm]{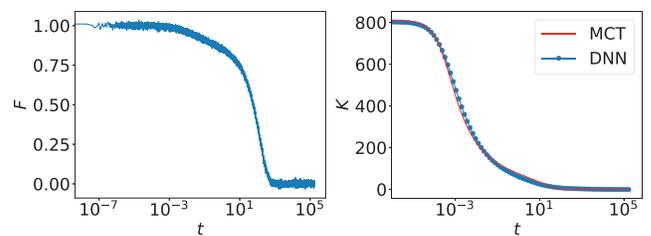}
\caption{The performance of a network trained on phenomenological data, applied to hard sphere MCT in the supercooled regime. Left panel: the input noisy $F(t)$ calculated from hard sphere MCT. Note this curve exhibits the two step relaxation that is typical of supercooled liquids. Right panel: the true memory kernel (`MCT') and the kernel predicted by the network (`DNN').%On the left is the input noisy $F(t)$ calculated by hard sphere MCT and exhibiting the two step relaxation that is typical of supercooled liquids. On the right is the true kernel (`MCT') and the kernel predicted by the network (`DNN').
}
\label{fig:phenom_MCT_results}
\end{figure}

Having trained the DNN on this phenomenological dataset, we then apply it to a noisy $F(t)$ of hard sphere MCT in the supercooled regime, where the dynamics exhibit a typical two step relaxation. The result is shown in Fig.\ \ref{fig:phenom_MCT_results}, where the DNN makes a very accurate prediction of the true MCT kernel despite never having seen MCT data. Fig.\ \ref{fig:phenom_MCT_results} demonstrates the ability of our deep learning approach to generalise beyond phenomenological training data to physically realistic unseen examples.

\section{\label{sec:conclusion} Conclusion}
In this work we develop a novel deep learning method for measuring the memory kernel of GLEs. Our method is generally applicable to any GLE system where a training set can be constructed. We demonstrate its effectiveness on MCT hard sphere data, and show that the DNN measures highly accurate kernels on unseen inputs, with particular robustness to noise. This is in stark contrast with existing Laplace inversion methods that are highly sensitive to noise. Furthermore, the DNN method generalises well to a system different from that used for training, and can even be used to make accurate predictions without training data generated from rigorous theory.

Our DNN method has several attractive features. It is computationally efficient as the time to generate a training set and train the network can be significantly less than the corresponding simulation time required by conventional methods, and making predictions with a trained network requires negligible computational resources. It also makes no distinction between short or long memory effects. This is in contrast to existing iterative techniques where errors accumulate over long times, posing a particular problem with long-lived glassy kernels. The ability of the DNN to map noisy single trajectory inputs to clean, ensemble averaged outputs is particularly powerful, allowing it to be used in situations where it is difficult to measure multiple trajectories for averaging, e.g.\ costly simulations or experiments. 

 As is the case with any machine learning technique, the network performs less well when presented with data that is dissimilar to that used during training, as seen with the low-temperature WCA data with low noise in Fig.\ \ref{fig:WCA_results}$\mathbf{(d)}$. However, this shortcoming can be addressed by including more diverse examples in the training set. As such, we present our code as a pipeline for training networks on data of the user's choosing, as well as our own trained network for the case of hard spheres. Giving users the tools to train networks on their specific problem will result in significantly smaller, yet better performing, networks than attempting to train a general-purpose GLE kernel extractor. The success with which neural networks can learn the highly non-trivial mapping between $F(t)$ and $K$ suggests that deep learning techniques should be considered for a diverse range of inverse problems where a training set can be generated by solving the simpler forward problem. \\

\section{Acknowledgements}
It is a pleasure to thank Sonja Georgievska, Meiert Grootes, and Jisk Attema of the Netherlands eScience Center for many valuable discussions in the context of the Small-Scale Initiative on Machine Learning. We acknowledge the Dutch Research Council (NWO) for financial support through a START-UP grant (MKW, VED, and LMCJ) and Vidi grant (IP and LMCJ). 

\appendix
\section{Hyperparameter search}\label{sec:appendix_paramsearch}
A hyperparameter search was performed over the L2 regularisation parameter, $\lambda$, the batch size, $B$, and width factor, $\omega$. For each combination of parameters, 5 networks with different initial conditions were trained. The initial condition was set by drawing weights and biases from a random uniform distribution $U(-\sqrt{k}, \sqrt{k})$, where $1/k$ is the number of input features to the layer. The hyperparameter values in the search were \\

\begin{itemize}
\item $\lambda$: 0.1, 0.05, 0.01, 0.001, 0
\item $B$: 300, 2500, 10000
\item $\omega$: 2, 4, 8,
\end{itemize}

resulting in a total of $5\times3\times3\times5=225$ networks. The depth of the network (6 hidden layers) is somewhat arbitrary, and was chosen by balancing the greater expressivity of deep networks with the increase in training time as the network gets larger. The Adam optimizer parameters, referred to as $\beta_1$ and $\beta_2$ in \cite{Kingma2015}, were kept as their default values of 0.9 and 0.999 respectively. The learning rate, $\mu_{\text{LR}}$, was set to $10^{-3}$. It would have been preferable to include the hyperparameters $\beta_1$, $\beta_2$, and $\mu_{\text{LR}}$ in our grid search, however this was computationally infeasible. 

\section{Simulation details}\label{sec:appendix_A}
Simulations were performed of a set of $N=2000$ Brownian particles in 3D, with periodic boundary conditions and a number density $\rho=0.95$. The position of particle $i$, $\mathbf{r}_i$, obeys the overdamped Langevin equation,
\begin{equation}\label{eq:brownianmotion}
    \dot{\mathbf{r}}_i = \zeta^{-1} \mathbf{F}_i  + \boldsymbol{\xi}_i(t),
\end{equation}
where $\zeta=1$ is a friction coefficient, $\mathbf{F}_i$ is a force acting on particle $i$ due to the inter-particle potential, and $\boldsymbol{\xi}_i$ is a random noise term obeying $\langle\boldsymbol{\xi}_i(t)\rangle=\mathbf{0}$, and $\langle\boldsymbol{\xi}_i(t)\cdot\boldsymbol{\xi}_j(t')\rangle=6D_0\delta_{ij}\delta(t-t')$, where $D_0 = k_BT/\zeta$ is the diffusion constant, $k_B$ the Boltzmann constant, and $T$ the temperature. The interaction force comes from the Weeks-Chandler-Andersen potential,
\begin{equation}
    U(r) = 4\epsilon\left[\left(\frac{\sigma}{r}\right)^{12} - \left(\frac{\sigma}{r}\right)^6\right] + \epsilon,
\end{equation}
where $r$ is the inter-particle distance. We use Lennard-Jones units such that $\sigma=1$ is the particle diameter, and $\epsilon=1$ a parameter determining the strength of the interaction. A dimensionless temperature can be defined as $T^\ast = k_BT/\epsilon$. $U(r)$ is truncated such that the potential is purely repulsive. The simulations were performed with the LAMMPS molecular dynamics software \cite{Thompson2022} with timestep $\Delta t = 10^{-5}$. The system is left to evolve for $10^7$ timesteps to equilibrate, then measurements are taken over a further $10^7$ timesteps.

\section{Avoiding overfitting}\label{sec:appendix_overfitting}
\begin{figure}[h]
\includegraphics[width=8.5cm]{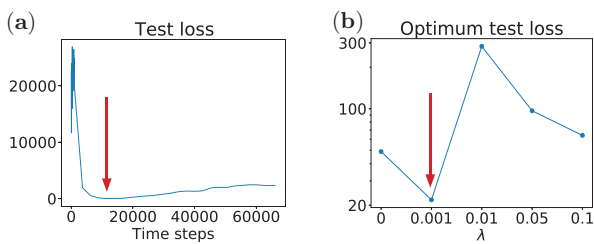}
\caption{$(\mathbf{a})$ The test loss of the DNN during training. Beyond a certain time point the system begins to overfit and the test loss increases. Consequently, we use the state of the network at the minimum test loss, as indicated by the red arrow. Note that the x axis is in time steps, not epochs. 1 epoch = $P/B$ timesteps. $(\mathbf{b})$ The minimum test loss achieved by networks with varying $\lambda$ (and $\omega=8$, batch size=2500). We select the $\lambda$ that results in the minimum test loss, as indicated by the red arrow.}
\label{fig:overfitting}
\end{figure}

Overfitting is a common problem in deep learning, particularly when the size of the network is much larger than the size of the dataset, as is the case in this work. Several methods have been suggested for avoiding overfitting, including but not limited to regularisation, early stopping, and the use of dropout layers. We use all three, and also perform a parameter search over the regularisation parameter $\lambda$ to find the optimum level of regularisation. In Fig.\ \ref{fig:overfitting}$(\mathbf{a})$ we demonstrate our use of early stopping, where we use the state of the network when it achieves the minimum test loss, not at the end of training. In Fig.\ \ref{fig:overfitting} $(\mathbf{b})$ we show the effect of varying $\lambda$. Due to limited computational resources we did not perform a parameter search over the dropout probability, instead setting it to 0.5 for all networks.

\section{Exploring unseen phase space}\label{sec:appendix_B}
\begin{figure}[h]
\includegraphics[width=9cm]{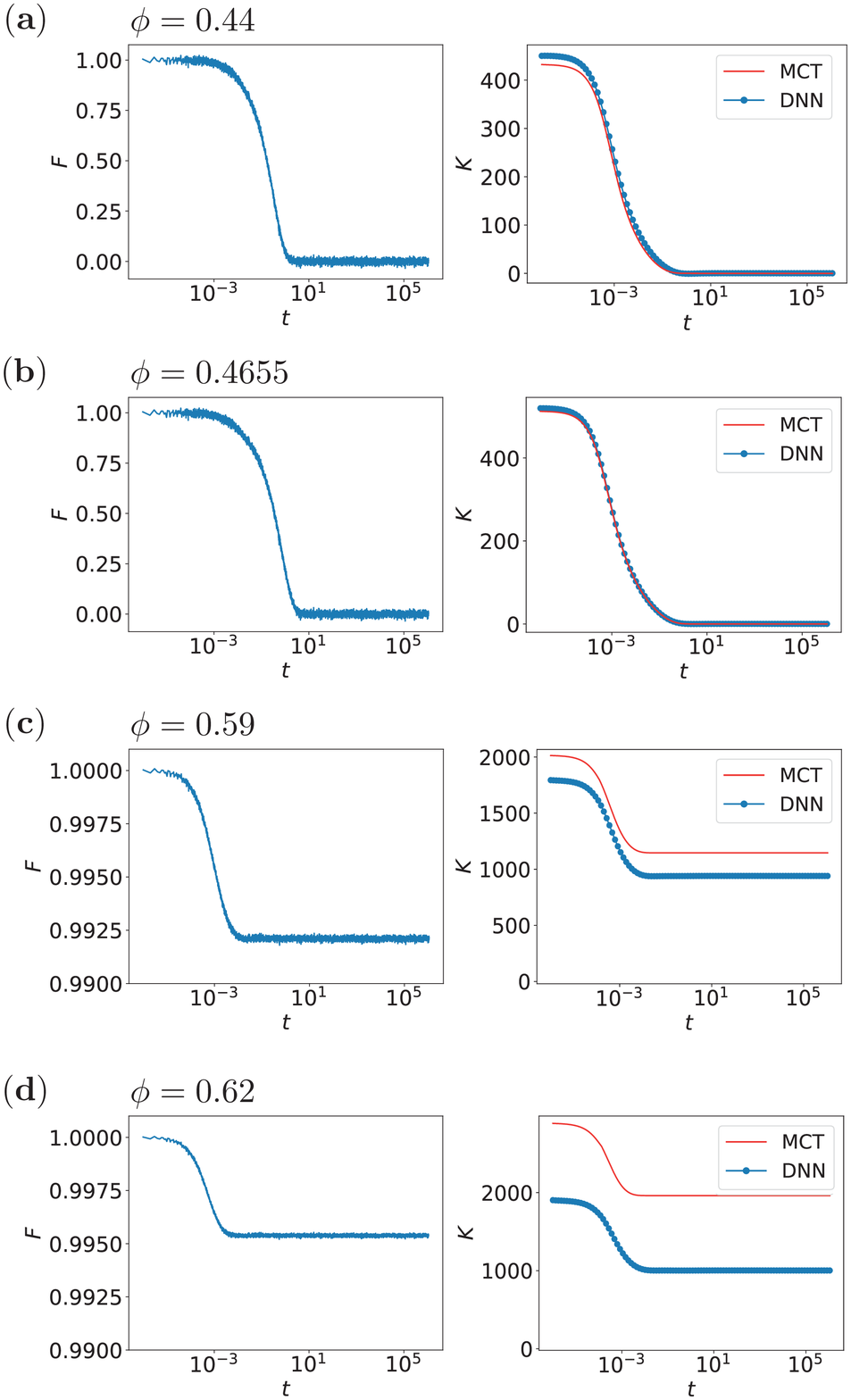}
\caption{$F(t)$ curves and the corresponding kernels, ground truth given by MCT and approximated by the network, at volume fractions not in the training set. $\mathbf{(a)}$ Below the lowest $\phi$ in the training set. $\mathbf{(b)}$ Between values in the training set. $\mathbf{(c)}$ and $\mathbf{(d)}$ Above the highest value in the training set.}
\label{fig:novel_phis}
\end{figure}

The network is trained on a set of $F(t)$ curves generated at volume fractions $\phi \in \{0.45, 0.451, 0.452..., 0.58 \}$, and subjected to multiple noise realisations. How does the network perform on volume fractions that are not in the training set? This question is addressed in Fig.\ \ref{fig:novel_phis}. Fig.\ \ref{fig:novel_phis}$\mathbf{(a)}$ shows the performance of the network at a volume fraction below the range of the training set. The left-hand panel is the input noisy $F(t)$, and the right-hand panel shows the measured $K$, which agrees with the true MCT kernel to a high degree of accuracy, with a small discrepancy at very short times. Fig.\ \ref{fig:novel_phis}$\mathbf{(b)}$ is at a volume fraction that is not in the training set, but is within the range of training $\phi$ values. In this case the network measures a highly accurate kernel, as it does at $\phi$ values in the training set. Fig.\  \ref{fig:novel_phis}$\mathbf{(c)}$ and $\mathbf{(d)}$ are at volume fractions above the upper end of the training set. Here the measured kernel is inaccurate. The reason for this can be seen in the $F(t)$ curves in the left-hand panels. At $\phi=0.59$ and $\phi=0.62$ we are deep in the glass phase, and the system decorrelates very little (i.e.\ the asymptote of $F(t)$ at long times is greater than 0.99 in both cases). As such, the dynamics become less and less sensitive to the exact form of the memory kernel. These plots demonstrate that the network can generalise well to regions of phase space beyond the training set, however care must be taken. For some parameter values (e.g.\ $\phi$ deep in the glass phase) the network is insufficiently sensitive to the weak relationship between $K$ and $F(t)$.

\section{Performance on soft spheres}\label{sec:WCA_comparison}
\begin{figure}[h]
\includegraphics[width=9cm]{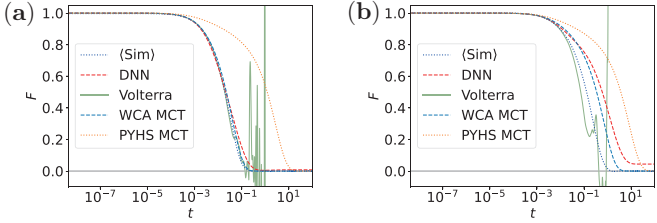}
\caption{A comparison of the dynamics from averaging many simulation trajectories ($\langle \text{Sim} \rangle$), solving the GLE with the kernel measured by the DNN method from a single simulation trajectory (`DNN'), the Volterra kernel (`Volterra'), the Weeks-Chandler-Andersen MCT kernel, and the Percus-Yevick Hard Sphere MCT kernel. All methods are presented at $T=3$ in $(\mathbf{a})$, and T=1 in $(\mathbf{b})$.}
\label{fig:WCA_comparison}
\end{figure}

Here we look in more detail at the performance of the DNN that has been trained on Percus-Yevick hard sphere MCT, and then applied to data from simulations of Weeks-Chandler-Andersen spheres. In Fig. \ref{fig:WCA_comparison} we plot the dynamics predicted by the DNN, along with the ensemble averaged dynamics of the simulations, the prediction of the Volterra method, and the dynamics as predicted by both HS and WCA MCT. In \ref{fig:WCA_comparison}$(\mathbf{a})$, the comparison is done at high temperature. In this plot, the DNN significantly outperforms HS MCT, despite being trained on this theory. This is because the network also uses as input the frequency term, $\Omega$, calculated from the simulation static structure factor. It is to be expected that at high temperature the Markovian dynamics (determined by $\Omega$) play a large role. In Fig. \ref{fig:WCA_comparison}$(\mathbf{b})$ the comparison is made at a lower temperature, hence the performance of the DNN is worse, though still better than HS MCT. In this example the WCA MCT prediction is also inaccurate, highlighting the approximate nature of the theory. These plots demonstrate the importance of knowing the limitations of the training set when applying machine learning methods. Better performance would be achieved by using a training set of either theoretical or phenomenological kernels that more accurately describe the low temperature behaviour of WCA particles.

\section{Generating a phenomenological dataset}\label{sec:phenom_appendix}
The phenomenological kernels are parameterised according to
\begin{align}
K(t) = \frac{a}{(1+b t^c)^d} + f e^{-(\frac{t}{10^g})^h}.
\end{align}
The training set was generated by combining three regions of parameter space: liquid, supercooled liquid, and glass. The liquid kernels used all combinations of parameter values in the following table:\\

\begin{tabular}{|l|l|}
  \hline
  a & $\{240, 275, 305 \}$ \\
  \hline
  b & $\{ 15, 12, 10 \}$ \\ 
  \hline
  c & $\{ 0.65, 0.6, 0.55 \}$ \\ 
  \hline
  d & $\{ 1.86, 1.73, 1.6 \}$ \\ 
  \hline
  f & $\{ 200, 240, 280 \}$ \\ 
  \hline
  g & $\{ -3.5, -3.25, -3.1 \}$ \\ 
  \hline
  h & $\{ 0.85, 0.83, 0.8 \}$ \\ 
  \hline
\end{tabular} \\

The supercooled kernels were from all combinations of the following parameters: \\

\begin{tabular}{|l|l|}
  \hline
  a & $\{660, 710, 760 \}$ \\
  \hline
  b & $\{ 12000, 30000, 100000 \}$ \\ 
  \hline
  c & $\{ 1.16, 1.26, 1.36 \}$ \\ 
  \hline
  d & $\{ 0.45, 0.35, 0.28 \}$ \\ 
  \hline
  f & $\{ 105, 90, 80 \}$ \\ 
  \hline
  g & $\{ 0.02, 1.0, 3.1 \}$ \\ 
  \hline
  h & $\{ 0.43, 0.49, 0.52 \}$ \\ 
  \hline
\end{tabular}\\

And finally the glass kernels were from all combinations of the following parameters:\\

\begin{tabular}{|l|l|}
  \hline
  a & $\{780, 795, 810 \}$ \\
  \hline
  b & $\{ 2800, 9000, 15000 \}$ \\ 
  \hline
  c & $\{ 0.98, 1.08, 1.16 \}$ \\ 
  \hline
  d & $\{ 0.5, 0.45, 0.4 \}$ \\ 
  \hline
  f & $\{ 5800, 2000, 140 \}$ \\ 
  \hline
  g & $\{ -300, -140, -8 \}$ \\ 
  \hline
  h & $\{ 0.002, -0.06, -0.114 \}$ \\ 
  \hline
\end{tabular} \\

These values were selected by first fitting the form of $K$ to the hard sphere MCT kernel at $\phi=0.45$, $0.515$ and $0.52$ (in the liquid, supercooled, and glass regimes respectively) then varying the parameters about their fitted values. Each kernel was used to solve the GLE, and the resulting $F(t)$ was subjected to four realisations of Gaussian noise. The (solution, kernel) pairs were then randomly shuffled and split 50:50 into training and testing sets.

\section{Volterra Method}\label{sec:volterra_appendix}

Our goal is to find an expression for the irreducible memory kernel $K(k,t) = \frac{1}{Nk^2D_0}\left<R^*_\textbf{k} e^{\Omega^\dagger\QQ'\QQ t}R_\textbf{k}\right>$. Here, $R_\textbf{k}=\QQ\Omega^\dagger\rho_\textbf{k}$ is the fluctuating force, $\Omega^\dagger$ is the conjugate of the Smoluchowski operator, and $\QQ = 1-\PP$ is the projector on the space orthogonal to that spanned by the density modes, given by $\PP = \left.\rho_\textbf{k}\right>\left<\rho_\textbf{k}^*\rho_\textbf{k}\right>^{-1}\left<\rho_\textbf{k}^*\right.$. Additionally, we have defined a second projection $\PP' = \left.\rho_\textbf{k}\right>\left<\rho_\textbf{k}^*\Omega^\dagger\rho_\textbf{k}\right>^{-1}\left<\rho_\textbf{k}^*\Omega^\dagger\right.$, and its complement $\QQ'=1-\PP'$ (for more details see \cite{nagele1996} and \cite{Nagele1999}).

Since the evolution operator $e^{\Omega^\dagger\QQ'\QQ t}$ is hard to deal with, we apply the Dyson decomposition identity,
\begin{equation}
    e^{\Omega^\dagger\QQ'\QQ t}=e^{\Omega^\dagger t} - \int_0^t\mathrm{d}\tau e^{\Omega^\dagger \QQ'\QQ (t-\tau)}\Omega^\dagger (1-\QQ'\QQ) e^{\Omega^\dagger t},
\end{equation}
which yields a Volterra equation for the memory kernel
\begin{align}\label{eq:k1k3}
    K(k,t) = K_{\Omega^\dagger}(k,t) + \int_0^t d\tau K(k,t-\tau)W(k,\tau).
\end{align}
Here we have introduced the function
\begin{equation}\label{eq:KL}
    K_{\Omega^\dagger}(k,t)=(Nk^2D)^{-1}\left<R^*_\textbf{k} e^{\Omega^\dagger t}R_\textbf{k}\right>,
\end{equation}
and the correlation
\begin{equation}\label{eq:W}
    W(k,t) = (Nk^2D)^{-1}\left<\rho_\textbf{k}^*\Omega^\dagger e^{\Omega^\dagger t}R_\textbf{k}\right>,
\end{equation}
which both evolve with standard Brownian dynamics.

In order to solve the integral equation \eqref{eq:k1k3}, we first compute $K_{\Omega^\dagger}(k,t)$ and $W(k,t)$ at $k=7.0$ from the simulation trajectories. This we do by evaluating their definitions \eqref{eq:KL} and \eqref{eq:W}, averaging over 50 independently initialised simulation trajectories, a small number of time origins, and all allowed wave vectors in the range $k\in(7.0\pm0.1)$. We refer to averaging over both independent simulation trajectories and time origins as an ensemble average. For Fig.\ \ref{fig:WCA_results}$\mathbf{(a)}$ and $\mathbf{(c)}$, we omit the ensemble and time-origin average, in order to introduce more noise. 

Either the single trajectory or ensemble-averaged $K_{\Omega^\dagger}(k,t)$ and $W(k,t)$ are inserted in a discretised version of integral equation \eqref{eq:k1k3}. The memory kernel is subsequently found by solving the resulting system of equations \cite{Pihlajamaa2023}.

\bibliographystyle{apsrev4-1}
\bibliography{learning_K.bib}

\end{document}